\documentclass[fleqn,12pt,twoside]{article}
\usepackage{amsmath}
\usepackage[headings]{espcrc1}

\readRCS
$Id: espcrc1.tex,v 1.2 2004/02/24 11:22:11 spepping Exp $
\usepackage{graphicx}
\usepackage[figuresright]{rotating}

\title{From Color Fields to Quark Gluon Plasma}

\author{Rainer~J.~Fries\address[MCSD]{School of Physics and Astronomy, University of Minnesota,\\ Minneapolis MN 55455, USA}%
, Joseph I.~Kapusta\addressmark[MCSD] and Yang Li\addressmark[MCSD]}
       
\runtitle{From Color Fields to QGP}
\runauthor{R.~J.~Fries}

\begin{document}

\maketitle

\begin{abstract}
We discuss a model for the energy distribution and the early space-time 
evolution of a heavy ion collision. We estimate the gluon field
generated in the wake of hard processes and through primordial fluctuations
of the color charges in the nuclei. Without specifying the dynamical mechanism
of thermalization we calculate the energy momentum tensor of the following
plasma phase. The results of this model can be used as initial conditions
for a further hydrodynamic evolution.
\end{abstract}

\section{INTRODUCTION}

Our knowledge about the early phase of a heavy ion collision, 
is still incomplete. Hydrodynamic calculations compared with data collected 
at the Relativistic Heavy Ion Collider (RHIC) suggest a rather early 
thermalization time $\tau_f < 1$ fm/$c$, at which a deconfined quark gluon 
plasma is created \cite{Adcox:2004mh}.
After the time $\tau_f$ ideal hydrodynamics delivers a successful 
description of the dynamics of the fireball. The initial conditions 
for the hydrodynamic evolution however remain obscure and are often 
subject to an educated guess.


Here we propose a simple model for the initial state that focuses on
two important questions: how much energy is deposited in the fireball
between the receding nuclei, and how is this energy distributed in space-time
at an early time $\tau_f$? 
The model uses three simple ingredients: an initial distribution of color 
sources in the colliding nuclei, (ii) the evolution of a classical 
McLerran-Venugopalan gluon field, (iii) energy-momentum conservation.

\section{OUTLINE OF THE MODEL}


Let us choose a momentum scale $Q_0$ (which should be much larger than 
$\Lambda_{\rm QCD}$).
We divide gluon production into hard modes (hard processes with transverse 
momentum $p_T > Q_0$) and soft modes (modes of the gluon field with
$p_T < Q_0$). Hard modes can be rigorously treated within perturbative 
QCD (pQCD) and implemented in a parton cascade.
For the soft modes we take guidance from the McLerran-Venugopalan (MV) model 
and describe them by classical solutions to the Yang-Mills equation
  $[D_\mu,F^{\mu\nu}] = J^\nu$  and the 
continuity equation $ [D_\mu,J^\nu] = 0$ \cite{KoMLeWei:95}.
In the MV model the field is generated by local color fluctuations in 
the nuclei which are otherwise color neutral. 

In our case the classical $SU(3)$ current $J^\mu$ contains these 
primordial color distributions $\rho_1$ and $\rho_2$. They are infinitely 
thin sheets of charge propagating with the speed of light along the 
light cone from 
$t=-\infty$ to $t=0$. Hard processes take place on very short time 
scales $\tau \sim 1/p_T$. They happen almost instantaneous for the soft fields.
For consistency color charges induced by hard processes have to be 
taken into account as additional sources of the soft gluon field, which are 
switched on at $t=0$. 
This is similar to a color ionization process: a parton with a 
certain color is participating in a hard process, leaving a hole 
with the corresponding anticolor. The parton can instead be found in the 
forward light cone.
Therefore $J^\mu$ contains modified color distributions $\rho_1'$ and $\rho_2'$
propagating on the light cone from $t=0$ to $t=+\infty$ and a charge $\rho_3$
in the forward light cone.

\begin{figure}[htb]
\begin{minipage}[t]{70mm} 
\includegraphics[width=6cm]{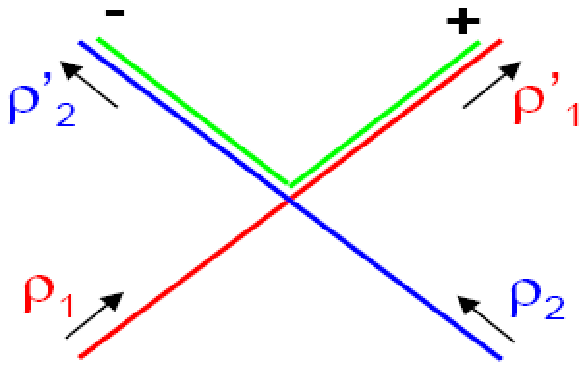}
\caption{The propagation of the color charges defined above 
on the lightcone.}
\label{fig:1}
\end{minipage}
\hspace{\fill}
\begin{minipage}[t]{85mm} \vglue 0pt \vspace{-3.5cm}
\includegraphics[width=8.5cm]{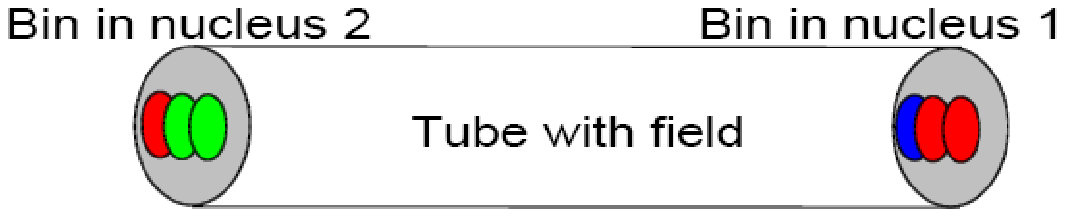}
\caption{Charges in opposite bins in both nuclei and the tube connecting them
which contains
the gluon field generated by the interaction. The partons in both bins
are represented by effective color charges $T_1$ and $T_2$.}
\label{fig:2}
\end{minipage}
\end{figure}

Here we shall not discuss the most general case, but demonstrate the basic
features of the model for the simple case of a boost invariant current.
In an axial gauge, defined by the condition $x^+ A^- + x^- A^+ = 0$ for the 
gauge potential $A$, we consider the current
$J^i (x) = 0$ ($i=1,2$) and
\begin{equation}
  \label{eq:current}
 J^\pm (x) = \delta(x^\mp) \left[\rho_{1,2}(\vec x_\perp) \pm \Theta(x^\pm) 
  \hat \rho(\vec x_\perp)\right],
\end{equation}
with $\rho'_{1,2} = \rho_{1,2} \pm \hat \rho$, see Fig.\ \ref{fig:1}.
We solve the Yang-Mills equations for small times $\tau$.
Causality ensures that only charges with a transverse distance less than
$c \tau$ can interact during this time. This suggests to 
discretize the charge distributions in the transverse plane. We use a 
rectangular grid of size $a$ for both nuclei. In each bin the classical 
charge with an effective size $1/\mu < a$ ($\mu\sim Q_0$) is put at the 
center. 
For $c\tau < a$ only the interaction of charges in the two nuclei which are 
opposite to each other (i.e.\ go through each other at the time of 
overlap, $t=0$) have to be taken into account.
The fireball between the receding nuclei is then described by a set of
parallel tubes of width $a$ with the bins of color charges at both ends 
(Fig.\ \ref{fig:2}).

In each bin in the nuclei the color charges of the partons are described by
an effective classical color charge. It was shown that this
classical charge is given by a random walk in 
$SU(3)$ space \cite{JeVe:04}. The probability distribution for the 
classical charge $T$ is given by
\begin{equation}
  w_N (|T|^2) = \left( \frac{N_c}{N\pi}\right)^4 e^{-N_c |T|^2/N}
\end{equation}
where the length $N$ of the random walk is determined by the number of quarks, 
antiquarks and gluons as $N = N_q + N_{\bar q} + (C_A/C_F) N_g$. 

The Yang-Mills equations can now be solved for each tube using
classical charges $T_1$ and $T_2$ as sources in the bins on either side.
We solve them to all orders in the coupling constant
$g$ using an expansion in powers of $\tau\mu$. 
At small times the dominating field component is the longitudinal 
chromoelectric field $F^{+-}$ in the tube. It has a finite value for
$\tau=0$, while the transverse chromoelectric and magnetic fields
$F^{i\pm}$ vanish initially and then increase linearly at early times.
The longitudinal chromomagnetic field $F^{ij}$ is zero at least up to order
$\tau^3$. We have solved the field equations up to order $\tau^3$ for the
longitudinal fields and up to order $\tau^4$ for the transverse fields.

\begin{figure}[htb]
\begin{minipage}[t]{75mm} \vglue 0pt \vspace{-0.5cm}
\includegraphics[width=7.5cm]{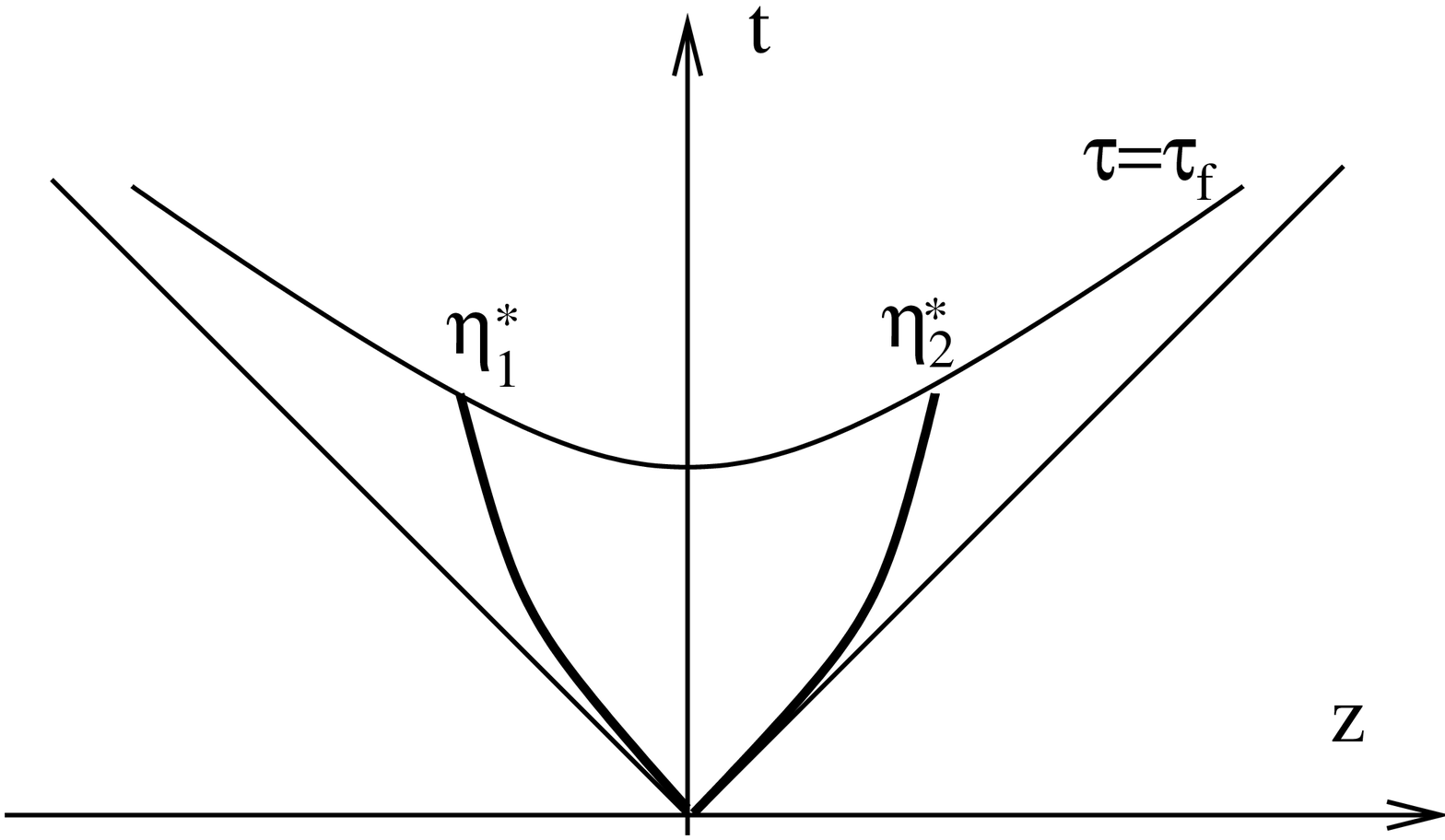}
\caption{The longitudinal structure of a tube in the fireball. The
partons in the bins to the left and right are decelerated. They reach final
rapidities $\eta_1^*$ and $\eta_2^*$ at the thermalization time $\tau_f$.}
\label{fig:3}
\end{minipage}
\hspace{\fill}
\begin{minipage}[t]{75mm} \vglue 0pt \vspace{-1.0cm}
\includegraphics[width=7.0cm]{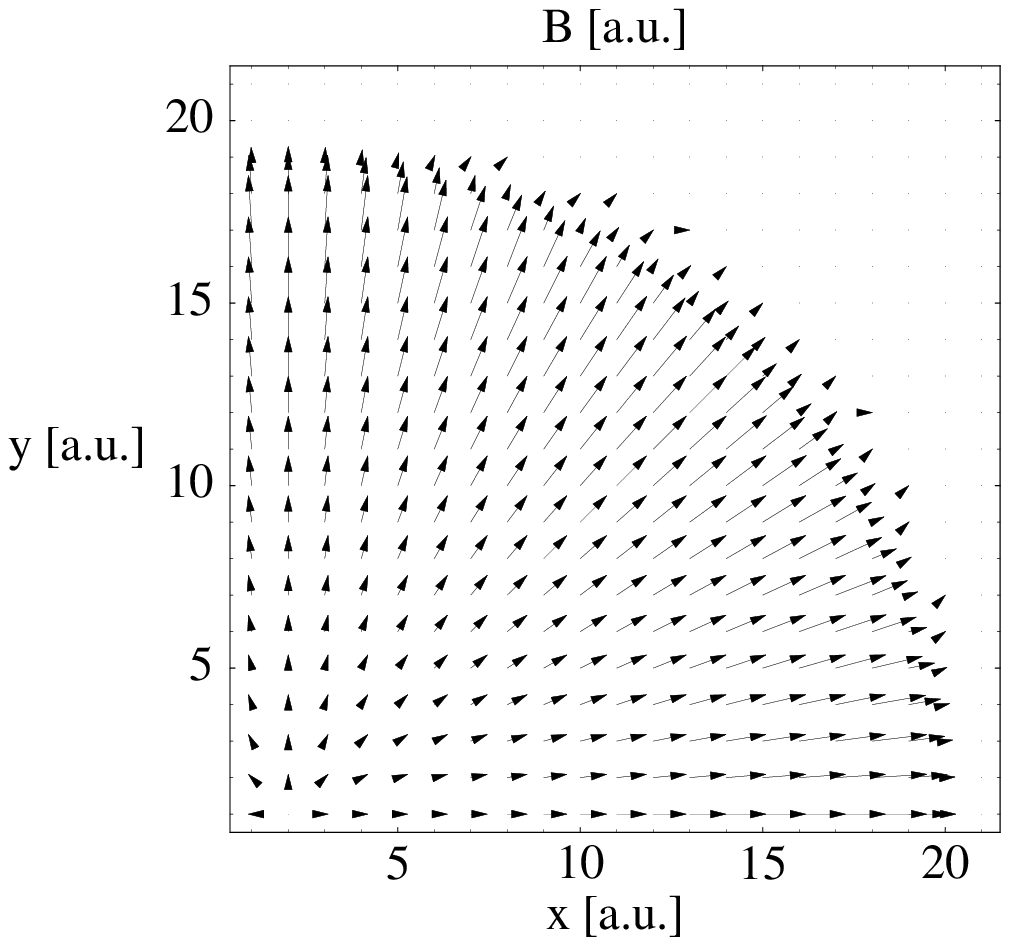}
\caption{A sketch of the 2-vector $\vec{B}$ in the transverse 
plane which determines the radial flow in the plasma phase.}
\label{fig:4}
\end{minipage}
\end{figure}

Thus at early times the field between the receding nuclei is analogous
to a capacitor as discussed in \cite{MiKa:01}. The longitudinal chromoelectric
field will slow down the charges. The trajectory of each bin of partons
for $t>0$ can be calculated to lowest order by assuming that it initially 
travels with the rapidity $y_0$ of the beam. We can follow the
trajectory to where it intersects the hypersurface $\tau=\tau_f$ at  
space-time rapidities $\eta_1^*$ and $\eta_2^*$. These values limit the 
longitudinal extent of the tube with the gluon field or plasma 
(see Fig.\ \ref{fig:3}).

At times $c \tau \approx a$ the fields of neighboring tubes start to overlap.
We add them coherently, and calculate the resulting energy-momentum tensor 
$T_f^{\mu\nu}$ from the gluon field. It turns out that the energy-momentum
tensor can be parameterized in the following way:
\begin{equation}
  T^{\mu\nu}_{\mathrm f} = 
  \begin{pmatrix} 
    A+C \cosh 2\eta & B_1 \cosh\eta & B_2 \cosh\eta & C \sinh 2\eta \\
    B_1 \cosh\eta & A+D & E & B_1 \sinh\eta \\
    B_2 \cosh\eta & E & A-D & B_2 \sinh\eta \\
    C \sinh 2\eta & B_1 \sinh\eta & B_2 \sinh\eta & -A +C \cosh 2\eta
  \end{pmatrix}
  \label{eq:genfieldtmunu}
\end{equation}
where $A$, $\vec B=(B_1,B_2)$, $C$, $D$ and $E$ are functions of time $\tau$
and the transverse coordinate $x_\perp$ but independent of space-time
rapidity $\eta$. At early times they behave like $A \sim \tau^0$, $\vec B \sim
\tau^1$ and $C\sim \tau^2$ respectively, while the functions $D$ and $E$ 
will not play any further role. 
Using the expansion in $\tau$ for the gluon field as
above we can ensure that $\partial_\mu T_f^{\mu\nu} = 0 +\mathcal{O}(\tau^2)$.

We can now match this field tensor with the tensor of the plasma phase,
$T_{\rm pl}^{\mu\nu} = (\epsilon+p) u^\mu u^\nu - pg^{\mu\nu}$ characterized by
a local energy density $\epsilon$, pressure $p$ and a 4-velocity 
$u = \gamma(1,\vec{v}_\perp,v_L)$ with longitudinal and transverse flow
$v_L$ and $\vec{v}_\perp$. Without detailed knowledge of the decay and
thermalization of the gluon field we can still make use of the fact that
$\partial_\mu T^{\mu\nu} = 0$ where $T_{\mu\nu}$ is the total energy momentum 
tensor of the fireball which is equal to $T_f^{\mu\nu}$ up to some
time before the thermalization time $\tau_f$ and equal to $T_{\rm pl}^{\mu\nu}$
after that. 

Here we discuss the case of an instantaneous matching for 
which $T^{\mu\nu} = \Theta(\tau_0-\tau) T_f^{\mu\nu} + \Theta(\tau-\tau_0)
T_{\rm pl}^{\mu\nu}$. Energy-momentum conservation implies a set of 
equations for the plasma:
\begin{equation}  \vec{v}_\perp = \frac{1}{\cosh\eta} \frac{\vec{B}}{A+C+p} 
  ,\quad v_L = \tanh\eta,\quad
  \epsilon + p = (A+C) \left( 1-\frac{\vec{B}^2}{(A+C+p)^2} \right)
\end{equation}
These equations fix transverse flow to be in the direction
of $\vec{B} \propto T_f^{0i}$. We also note that the longitudinal flow
leads to a boost-invariant scenario where rapidity $y$ is equal to $\eta$
up to the limiting values $\eta^*_{1,2}$. Of course this is a consequence of 
the boost-invariant charge distribution in this example. 
We need another equation to unambiguously determine the magnitude of 
$v_\perp$, $p$ and $\epsilon$. This can be the equation of state which
is used for the hydrodynamic evolution for $\tau > \tau_f$. E.g.\ an ideal
gas equation of state would lead to 
$\epsilon = -(A+C)+\sqrt{4(A+C)^2 -3 B^2}$.

\section{SUMMARY}

To summarize, we introduce a model for the early phase of a heavy ion
collision that can be used to estimate how much energy is deposited 
by the production of gluon fields and how the energy is distributed once
the system thermalizes.
In this model it is natural to have transverse flow and also
elliptic flow already at the beginning of the plasma phase. Flow 
originates from the transverse energy flow $T_f^{0i}$ of the early gluon 
phase, see Fig.\ \ref{fig:4}. Preliminary results have been obtained 
using a boost invariant charge distribution in the nuclei. 
Some of the details remain to be carried out in more detail.
In particular we aim for a consistent implementation of hard processes 
which minimizes the dependence of the final results on the hard scale $Q_0$.
Hydrodynamic simulations using our results are in preparation.

\end{document}